\documentclass[final]{raa}
\usepackage{graphicx,times}
\usepackage{natbib}
\usepackage{amssymb,amsmath}
\bibpunct{(}{)}{;}{a}{}{,}

\usepackage[a4paper=true,dvipdfm=true,pagebackref=true]{hyperref}
\hypersetup{pdftitle = The title of my PDF, pdfauthor = My name, pdfsubject= The subject, pdfkeywords = keyword1 keyword2 keyword3}
\hypersetup{colorlinks = true, linkcolor = green, anchorcolor = red, citecolor = blue, filecolor = red, pagecolor = red, urlcolor = red}

\begin{document}

   \title{Singlet pairing gaps of neutrons and protons in hyperonic neutron stars}

 \volnopage{ {\bf 201x} Vol.\ {\bf X} No. {\bf XX}, 000--000}
   \setcounter{page}{1}

   \author{Yan Xu\inst{1}, Cheng-Zhi Liu\inst{1}, Cun-Bo Fan\inst{1}, Xing-Wei~Han
      \inst{1},  Xiao-Jun Zhang\inst{1}, Ming-Feng Zhu\inst{2}, Hong-Yan Wang\inst{3}, Guang-Zhou Liu\inst{2}
   }

   \institute{ Changchun Observatory, National Astronomical Observatories,
Chinese Academy of Sciences, Changchun 130117, China;\\{\it Corresponding Author. C.Z.Liu, lcz@cho.ac.cn; Y.Xu, xuy@cho.ac.cn}\\
        \and College of Physics, Jilin University, Changchun 130012, China\\
    \and College of Physics, Beihua University, Jilin 132013, China\\
\vs \no
   {\small Received 2014 May 26; accepted 2014 July 28}
}

\abstract{The $^{1}S_{0}$ nucleonic superfluids are investigated
within the relativistic mean-field model and
Bardeen-Cooper-Schrieffer theory in hyperonic neutron stars. The
$^{1}S_{0}$ pairing gaps of neutrons and protons are calculated
based on the Reid soft-core interaction as the nucleon-nucleon
interaction. We have studied particularly the influence of hyperons
degrees of freedom on the $^{1}S_{0}$ nucleonic pairing gap in
neutron star matter. It is found that the appearance of hyperons has
little impact on baryonic density range and size for the $^{1}S_{0}$
neutronic pairing gap,  the $^{1}S_{0}$ protonic pairing gap also
decreases slightly in this region $\rho_B=0.0-0.393$ fm$^{-3}$.
However, if baryonic density becomes greater than 0.393 fm${^{-3}}$,
the $^{1}S_{0}$ protonic pairing gap obviously increases. In
addition, the protonic superfluid range is obviously enlarged due to
the presence of hyperons. In our results, the hyperons change the
$^{1}S_{0}$ protonic pairing gap which must change the cooling
properties of neutron stars.  }

   \authorrunning{Y. Xu et al. }            
   \titlerunning{$^{1}S_{0}$ Nucleon Pairing Gap}  
   \maketitle

%
\section{Introduction}           
\label{sect:intro}

In resent years, neutron stars(NSs) have been becoming one of the
hottest scientific problems in the domain of astrophysics. The
reasoning is that the nucleonic energy gap and corresponding
superfluidity(SF) critical temperature can greatly affect the
neutrino emission which dominated about $10^{5}$-$10^{6}$ years of
NS cooling
phases(\citealt{Zuo+etal+2004,Zuo+Lombardo+2010,Gao+etal+2011,Tanigawa+etal+2004,Kaminker+etal+2002,Xu+etal+2013,Xu+etal+2012a,raa1,raa2}).
Neutrons, protons in the interior of NSs can become the superfluid
states due to the attraction between two neutrons or protons.
Neutrons in NS crust probably form $^{1}S_{0}$ pairing and in NS
core mainly form $^{3}P_{2}$ pairing. Protons in NS core can suffer
$^{1}S_{0}$ pairing which appear in NS matter with supranuclear
density, such density region is related largely to the direct Urca
processes on
nucleons(\citealt{Yakovlev+etal+1999,Shternin+etal+2011,Chen+etal+2006}).
And it is well known that the direct Urca processes on nucleons
produce the most powerful neutrino energy
losses(\citealt{Haensel+Gnedin+1994,Gnedin+etal+1994,Yakovlev+etal+2008,Xu+etal+2014}).
Thus nucleonic superfluids must affect NS cooling.

The $^{1}S_{0}$ nucleonic pairing gap has been considered using
different model potentials of the nucleon-nucleon(NN) interaction.
These theoretic calculations based on qualitative models give
similar ranges for the presence of the $^{1}S_{0}$ nucleonic
pairing. Nevertheless, due to many uncertain factors about the NN
interaction such as the non-direct observational data in extreme
conditions, approximations used in the calculations, it can not get
the accurate results for the pairing gap and estimate the superfluid
quantitative influence on NSs. Nowdays, many of relativistic models
are drawing attention in studies on NSs because they are
particularly suited for describing NSs according to the special
relativity. The most common among them is the relativistic mean
field(RMF) theory which could give very successful descriptions in
nuclear matter and finite nuclei researches
(\citealt{Glendenning+1985}). In 1993, to adapt to the effective
$\Lambda$$\Lambda$ interaction $\Delta B_{\Lambda\Lambda}\sim5$MeV
inferred from the earlier measurement and avoid the high-density
instability in numerical calculation, Schaffner et al extended the
standard RMF model by adding strange mesons $\sigma ^{*}$ and
$\phi$(\citealt{Schaffner+etal+1993}). That is, baryons interact by
exchanging the $\sigma$, $\omega$, $\rho$, $\sigma ^{*}$ and $\phi$
mesons. While the recent measurement suggests that $\Delta
B_{\Lambda\Lambda}$ should be $1.01\pm
0.20^{+0.18}_{-0.11}$MeV(\citealt{Bednarek+Manka+2005,Yang+Shen+2008,Wang+Shen+2010,Xu+etal+2012b}).
In addition, when baryonic density is lower than
$2\rho_{0}$($\rho_{0}$ is the saturation density of nuclear matter),
NSs are generally made up of only neutrons, protons and leptons. Yet
if baryonic density exceeds about $2\rho_{0}$, hyperons as new
degrees of freedom will appear in NSs cores. They must result in the
equation of state(EOS) changing, such the nucleonic Fermi momenta
and single particle energies tend to change. So the hyperonic
appearance would affect the nucleonic SF.

The content of the paper is arranged in this way. The NS properties
and $^{1}S_{0}$ nucleonic pairing gap are described using RMF and
Bardeen-Cooper-Schrieffer(BCS) theories in Sec.2. The numerical
results are displayed in Sec.3. The summary is presented in Sec.4.
$\Sigma$ hyperons are ruled out due to the debatable $\Sigma$
potential at $\rho_{0}$ in nuclear matter
(\citealt{Batty+etal+1994}). The paper is primarily focused upon the
influence of hyperons on the $^{1}S_{0}$ nucleonic pairing gap.
\section{The models}
\label{sect:Obs}

Baryonic interactions are by exchanging of $\sigma$, $\omega$,
$\rho$, $\sigma^{*}$, $\phi$ mesons in RMF approach. In this paper,
n, p, $\Lambda$ and $\Xi$ baryons are considered in NSs. The
contribution of the first three mesons to the lagrangian is
(\citealt{Glendenning+1985}),
\begin{multline}
\hspace{-5mm}
L=\sum_B\overline{\psi}_B[i\gamma_\mu\partial^\mu-(M_B-g_{\sigma
B}\sigma) -g_{\rho
B}\gamma_\mu{\boldsymbol{\tau}}\cdot{\boldsymbol{\rho}}^\mu
-g_{\omega B}\gamma_\mu\omega^\mu
 ]\psi_B+\frac{1}{2}(\partial_\mu\sigma\partial^\mu\sigma-m_\sigma^2\sigma^2)-U(\sigma)
   +\frac{1}{2}m_\omega^2
  \omega_\mu\omega^\mu\\+\frac{1}{4}c_3(\omega_\mu\omega^\mu)^2+
\frac{1}{2}m_\rho^2{\boldsymbol{\rho}}_\mu{\boldsymbol{\rho}}^\mu
-\frac{1}{4}F^{\mu v}F_{\mu v}-\frac{1}{4}G^{\mu v}G_{\mu
v}+\sum_l\overline{\psi}_l[i\gamma_\mu\partial^\mu-m_l]\psi_l,
\end{multline}
Where the field tensors of the vector mesons $\omega$ and $\rho$ are
marked as $F_{\mu v}$ and $G_{\mu v}$, respectively.
$U(\sigma)=\frac{1}{3}a \sigma ^{3}+\frac{1}{4}b \sigma ^4$. The
baryon species are marked as B. The contribution of strange mesons
$\sigma^{*}$, $\phi$ to the lagrangian is,
\begin{multline}
L^{YY}=\frac{1}{2}(\partial_v\sigma^*\partial^v\sigma^*-m^2_{\sigma^*}\sigma^{*2})-\sum_Bg_{\sigma^*B}\overline{\psi}_{B}\psi_{B}\sigma^*-
\sum_Bg_{\phi B}\overline{\psi}_{B}\gamma_\mu\psi_{B}\phi^\mu
-\frac{1}{4}S^{\mu v}S_{\mu
v}+\frac{1}{2}m^2_{\phi}\phi_\mu\phi^\mu.
\end{multline}
Here $\sigma^*$, $\phi$ aren't coupled with nucleons, they only
affect the hyperonic properties.

The five meson fields are considered as classical fields, and the
field operators are replaced with their expectation values in RMF
model(\citealt{Glendenning+1985,Bednarek+Manka+2005,Yang+Shen+2008,Wang+Shen+2010,Xu+etal+2012b}).
The meson field equations in NSs are as follows:
\begin{eqnarray}
\hspace{-2mm}
\sum_B g_{\sigma B}\rho_{SB}=m_\sigma^2\sigma+a\sigma^2+b\sigma^3,\\
\sum_B g_{\omega B}\rho_B=m_\omega^2\omega_0,\\
\sum_B g_{\rho B}\rho_{B}I_{3B} =m_{\rho}^2\rho_0,\\
\sum_B g_{\sigma^* B}\rho_{SB}=m_{\sigma^*}^2\sigma^*,\\
\sum_B g_{\phi B}\rho_B=m_\phi^2\phi_0.
\end{eqnarray}
Here $I_{3B}$ is the isospin projections of baryon species B.
$\rho_{S B}$ and $\rho_B$ mark baryonic scalar and vector densities,
respectively. They are,
\begin{eqnarray}
\rho_{SB}=\frac{1}{\pi^{2}}\int_0^{k_{F}}\frac{m_{B}^{*}}{\sqrt{k^{2}+m_{B}^{*2}}}k^{2}dk,
\nonumber
\\
\rho_{B}=\frac{k_{F}^{3}}{3\pi^2}.
\end{eqnarray}
Here $k_{F}$ is the baryonic Fermi momentum,
$m_{B}^*=m_{B}-g_{\sigma B}\sigma_{0}-g_{\sigma^* B}\sigma_{0}^*$ is
the baryonic effective mass.

A description of NS matter with uniform distribution is obtained
through the conditions of electrical neutrality and $\beta$
equilibrium. The electrical neutrality condition is
\begin{equation}
\rho_{p}=\rho_{\Xi^{-}}+\rho_{e}+\rho_{\mu}.
\end{equation}
The baryonic chemical potential is expressed by
\begin{equation}
 \mu_{B}=\mu_{n}-q_{B}\mu_{e},
\end{equation}
where $q_{B}$ is the baryonic electric charge number. Then the
$\beta$ equilibrium conditions are given by
\begin{eqnarray}
\mu_{n}=\mu_{p}+\mu_{e}, ~~~\mu_{\Xi^{-}}=\mu_{n}+\mu_{e} \nonumber
\\
\mu_{n}=\mu_{\Lambda}=\mu_{\Xi^{0}}, ~~~\mu_{e}=\mu_{\mu}.
\end{eqnarray}

The nucleonic single-particle energy in the model is
\begin{equation}
E_{N}(k)=\sqrt{k^2+{m_N^*}^2}+g_{\omega N} \omega_{0}+g_{\rho N}
\rho_{03} I_{3N}.
\end{equation}
The BCS gap equation is(\citealt{Zuo+etal+2004,Chen+etal+2006,
Xu+etal+2013}),
\begin{equation}
\Delta_{N}(k)=-\int{\frac{V_{NN}(k,k^{'})\Delta_{N}(k^{'})k^{'2}dk^{'}}{{4\pi^{2}}\sqrt{[(E_{N}(k^{'})-E_{N}(k_{F})]^{2}+\Delta_{N}^{2}(k^{'})}}}.
\end{equation}
The $^{1}S_{0}$ pairing gaps of neutrons and protons are calculated
based on the Reid soft-core(RSC)
interaction(\citealt{Nishizaki+etal+1991,Sprung+Banerjee+1971,Amundsen+stgaard+1985,Wambach+etal+1993}),
The $^{1}S_{0}$ channel interaction between two neutrons or protons
is
\begin{equation}
V_{NN}(k,k^{'})=4\pi\int{ r^{2}drj_{0}(kr)V_{NN}(r)j_{0}(k^{'}r)},
\end{equation}
where $V_{NN}(r)$ is the $^{1}S_{0}$ NN interaction potential in
coordinate space, $j_{0}(kr)$ is the zero order spherical Bessel
function.

The nucleonic critical temperature $T_{CN}$ of the $^{1}S_{0}$
pairing SF is(\citealt{Takatsuka+Tamagaki+2003}),
\begin{equation}
T_{CN}\doteq 0.66 \Delta_{N}(k_{F})\times10^{10}.
\end{equation}

According to the discussion of the RMF approach above, we can obtain
the EOS and NS composition as well as the nucleonic Fermi momenta
and single particle energies which are vitally important in the
research of the NN pairing gap.
\begin{table}
\bc
\begin{minipage}[]{100mm}
\caption[]{The TM1 set, the masses are in unit of MeV
(\citealt{Yang+Shen+2008})\label{tab1}}\end{minipage}
\setlength{\tabcolsep}{1pt} \small
 \begin{tabular}{ccccccccccc}
  \hline\noalign{\smallskip}
$m_{\sigma}$&$g_{\sigma N}$&$m_{\omega}$&$g_{\omega N}$&$m_{\rho}$&$g_{\rho N}$&$c_3$&$g_2$(fm$^{-1}$)&$g_3$&$m_{N}$&$m_{\phi}$\\
  \hline\noalign{\smallskip}
511.198&10.029&783.0&12.614&770.0&4.632&71.308&7.233 &0.618&983.0&1020.0\\
  \noalign{\smallskip}\hline
\end{tabular}
\ec
\end{table}

\begin{table}
\bc
\begin{minipage}[]{100mm}
\caption[]{The scalar coupling constants of hyperons, the masses are
in unit of MeV\label{tab2}}\end{minipage}
\setlength{\tabcolsep}{1pt}
 \begin{tabular}{ccccccc}
  \hline\noalign{\smallskip}
&$m_{\sigma^{*}}$&$m_{\phi^{*}}$&$g_{\sigma\Lambda}$&$g_{\sigma\Xi}$&$g_{\sigma^*\Lambda}$& $g_{\sigma^*\Xi}$\\
  \hline\noalign{\smallskip}
with$\sigma^*\phi$ &975.0&1020.0& 6.170& 3.202 &5.412 &11.516\\
  \noalign{\smallskip}\hline
\end{tabular}
\ec
\end{table}

\begin{table}
\bc
\begin{minipage}[]{100mm}
\caption[]{Baryonic density ranges for the $^{1}S_{0}$ nucleonic
pairing gap $\Delta_{N}(k_{F})$ at the Fermi surface in npe$\mu$ and
npHe$\mu$ matter\label{tab3}}\end{minipage}
\setlength{\tabcolsep}{1pt} \small
 \begin{tabular}{ccc}
  \hline\noalign{\smallskip}
 & npe$\mu$ & npHe$\mu$\\
  \hline\noalign{\smallskip}
$\Delta_{n}(k_{F})$  & $0.0 \leq \rho_{B}\leq 0.188$ & $0.0 \leq \rho_{B}\leq 0.191$\\
$\Delta_{p}(k_{F})$  & $0.0 \leq \rho_{B}\leq 0.509$ & $0.0 \leq \rho_{B}\leq 0.588$\\
  \noalign{\smallskip}\hline
\end{tabular}
\ec
\end{table}

\begin{figure}
   \centering
   \includegraphics[width=9.0cm, angle=0]{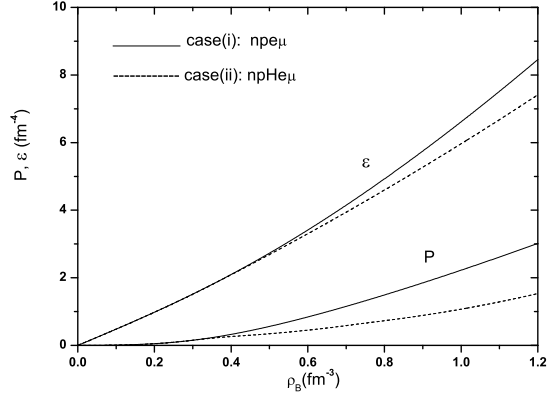}
   \caption{Pressure P and energy density $\varepsilon$ as functions of baryonic density $\rho_{B}$ in npe$\mu$ and npHe$\mu$ matter. }
   \label{Fig1}
   \end{figure}

\begin{figure}
   \centering
  \includegraphics[width=9cm, angle=0]{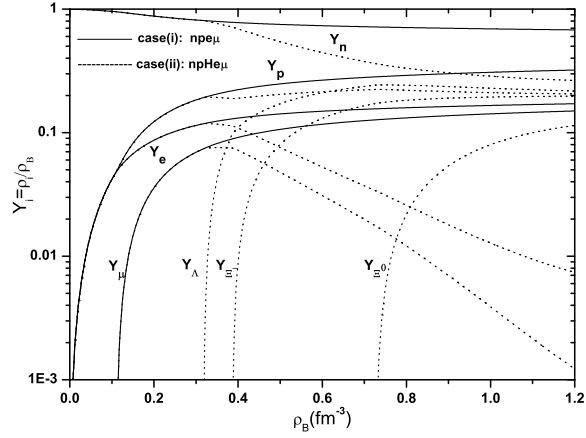}
   \caption{Composition of NSs as functions of baryonic density $\rho_{B}$.}
   \label{Fig2}
   \end{figure}

\begin{figure}
   \centering
  \includegraphics[width=9cm, angle=0]{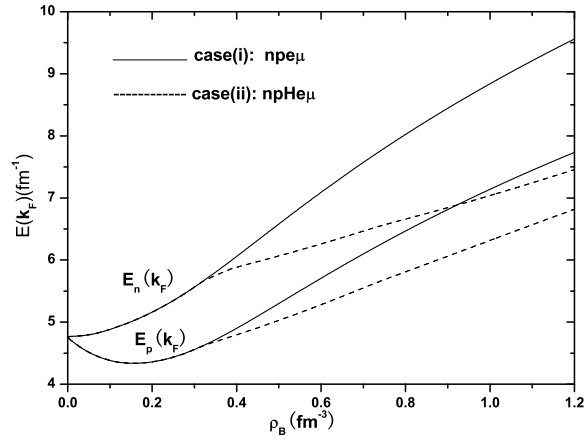}
   \caption{The nucleonic single particle energy $E_{N}(k_{F})$ at the Fermi surface vs baryonic density $\rho_{B}$ in npe$\mu$ and npHe$\mu$ matter.}
   \label{Fig3}
   \end{figure}

   \begin{figure}
   \centering
  \includegraphics[width=9cm, angle=0]{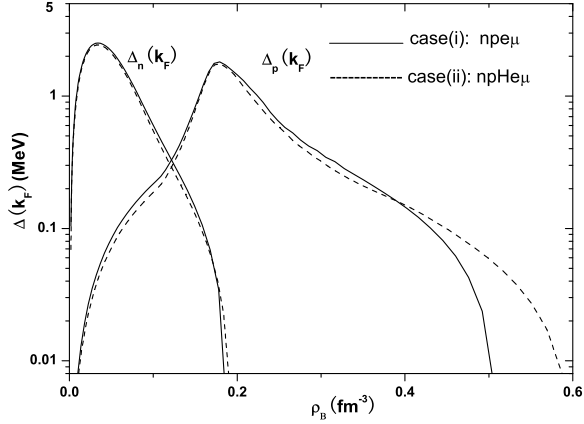}
   \caption{The $^{1}S_{0}$ nucleonic pairing gap $\Delta_{N}(k_{F})$ at the Fermi surface as functions of baryonic density $\rho_{B}$ in npe$\mu$ and npHe$\mu$ matter.}
   \label{Fig4}
   \end{figure}
\section{Discussion}
\label{sect:discussion}

According to the uncertainty of interior constitution of NSs, we
research NSs in both cases: (i) NS copnsition is n, p, e,
$\mu$(npe$\mu$), (ii) n, p, $\Lambda$, $\Xi^{0}$, $\Xi^{-}$, e,
$\mu$(npHe$\mu$). This work focuses on the influence of hyperons on
the $^{1}S_{0}$ nucleonic pairing gap in NSs. The appearance of
hyperons changes the EOS and NS composition as well as the
$^{1}S_{0}$ nucleonic pairing SF. It is widely accepted that the
$^{1}S_{0}$ nucleonic superfluids should be decided by pairing gap
$\Delta_{N}(k)$. Next, we'll show the numerical results for the
$^{1}S_{0}$ nucleonic pairing gaps in npe$\mu$ and npHe$\mu$ matter.
The NSs' properties are getted using a set of parameters displayed
in Table 1,2. We use $ U_\Lambda^N=-30$MeV, $U_\Sigma^N=+30$MeV,
$U_\Xi^N=-18$MeV, and $U_\Lambda^\Lambda=-5$MeV which are obtained
based upon the recent measurement $\Delta
B_{\Lambda\Lambda}\sim1.01\pm 0.20^{+0.18}_{-0.11}$MeV to decide the
hyperonic scalar coupling constants. We use $\frac{2}{3}g_{\omega
N}=g_{\omega\Lambda}= 2g_{\omega\Xi}$, $g_{\rho N}= g_{\rho\Xi}$,
$g_{\rho\Lambda}=0$,
$2g_{\phi\Lambda}=g_{\phi\Xi}=-\frac{2\sqrt{2}}{3}g_{\omega N}$ to
calculate the hyperonic vector coupling constants
(\citealt{Bednarek+Manka+2005,Yang+Shen+2008,Wang+Shen+2010,Xu+etal+2012b}).

As mentioned above, baryons interact by exchanging mesons. More
specifically, the attraction, repulsion and isospin interaction
between two baryons are supplied by $\sigma$, $\omega$ and $\rho$,
respectively. And the additional attraction and repulsion between
two hyperons are supplied by strange mesons $\sigma^*$ and $\phi$,
respectively. Fig.1 gives the EOS, namely, pressure $P$ and energy
density $\varepsilon$ as functions of baryonic density $\rho_{B}$ in
npe$\mu$ and npHe$\mu$ matter. As shown in Fig.1, one could see that
the pressure $P$ and energy density $\varepsilon$ remain unchanged
at lower densities in both cases. While with increasing of baryonic
density, the appearance of hyperons makes the pressure $P$ and
energy density $\varepsilon$ decrescent. That is, the EOS is
softened which will inevitably cause the NS's bulk property
changing. The crucial physical quantities for the $^{1}S_{0}$
nucleonic pairing gap are the nucleonic Fermi momenta and
single-particle energies. Fig.2 shows the numerical results of NS
composition as functions of baryonic density $\rho_B$ in npe$\mu$
and npHe$\mu$ matter. As shown in Fig.2, the threshold densities of
$\Lambda$, $\Xi^{-}$ and $\Xi^{0}$ hyperons are 0.320 fm$^{-3}$,
0.389 fm$^{-3}$ and 0.734 fm$^{-3}$, respectiely. One can also see
that nucleonic fractions are suppressed due to hyperons appearing in
NSs through the conditions of electrical neutrality and $\beta$
equilibrium(see Eqs.9-11). Therefore, according to Eq.8, we can see
that when hyperons appear in NSs, the individual Fermi momenta of
neutrons and protons are all much less than their values in npe$\mu$
matter. Fig.3 displays the nucleonic single particle energy
$E_{N}(k_{F})$ at the Fermi surface as functions of baryonic density
$\rho_B$ in npe$\mu$ and npHe$\mu$ matter. As Fig.3 shows, the
nucleonic single particle energies in npHe$\mu$ matter are obviously
less than their values in npe$\mu$ matter which is because the
reduction of the nucleonic Fermi momenta resulting in $E_{n}(k_{F})$
and $E_{p}(k_{F})$ all decreasing in npHe$\mu$ matter(see Eq.12).

So far, due to the uncertainty of NN interaction, the $^{1}S_{0}$
nucleonic pairing gap is also uncertain. In the work, we calculate
$\Delta_{N}(k)$ based on the RSC potential. The main concentration
of us is the hyperonic influences on the $^{1}S_{0}$ nucleonic
pairing gap. Fig.4 presents the $^{1}S_{0}$ nucleonic pairing gap at
the Fermi surface as functions of baryonic density $\rho_B$ in
npe$\mu$ and npHe$\mu$ matter. In Fig.4, one can see that the
$^{1}S_{0}$ neutronic pairing gap always exists in lower densities
region in both cases. The region of neutronic superfluids only
affects the NS's surface cooling. The $^{1}S_{0}$ protonic
superfluids can reach relatively high densities. The region of
protonic superfluids is closely related to the direct Urca processes
on nucleons which governs almost the cooling processes of NSs. In
addition, one can also see that the appearance of hyperons has
little impact on baryonic density range and size for the $^{1}S_{0}$
neutronic pairing gap $\Delta_{n}(k_{F})$ in Fig.4. This is because
the $^{1}S_{0}$ neutronic superfluids appear only in the region of
lower densities where hyperons have not appeared in NS matter. To
see clearly the influence of hyperons degrees of freedom on the
$^{1}S_{0}$ nucleonic pairing gap more intuitively, baryonic density
ranges for the $^{1}S_{0}$ nucleonic pairing gap $\Delta_{N}(k_{F})$
at the Fermi surface in npe$\mu$ and npHe$\mu$ matter are listed in
Table 3. As seen in Fig.4 and Table 3, when hyperons appear in NS
matter, the $^{1}S_{0}$ protonic pairing gap $\Delta_{p}(k_{F})$
decreases slightly in the region of $\rho_B=0.0-0.393$ fm$^{-3}$,
and increases obviously in the region of $\rho_B=0.393 - 0.588$
fm$^{-3}$ which is because the appearance of $\Lambda$ and $\Xi^{-}$
hyperons in NS core(see Fig.2 for details). The increase of the
$^{1}S_{0}$ protonic pairing gap must lead to the increase of the
protonic critical temperature $T_{CP}$(see Eq.15), then the neutrino
energy losses of the direct Urca on nucleon would be further
suppressed in the region of $\rho_B=0.393 - 0.588$ fm$^{-3}$. And
the range of the $^{1}S_{0}$ protonic SF is obviously enlarged due
to the presence of hyperons, which can achieve coverage or partial
coverage NSs cores. That is, If we don't consider the contributions
of the direct Urca processes on hyperons on NS cooling, the presence
of hyperons must decrease the cooling rate of NSs.

\section{Conclusion}
We study the effects of hyperons on the $^{1}S_{0}$ nucleonic SF by
adopting the RMF and BCS theories in NSs. The results indicate that
the appearance of hyperons has little influence on baryonic density
range and size for the $^{1}S_{0}$ neutronic SF. However, the
$^{1}S_{0}$ protonic pairing gap (and the $^{1}S_{0}$ protonic
critical temperature) in npHe$\mu$ matter is much larger than their
values in npe$\mu$ matter in the region of $\rho_B=0.393 - 0.588$
fm$^{-3}$. The baryonic density range of the $^{1}S_{0}$ protonic SF
is also enlarged from $\rho_B=0.0 - 0.509$ fm$^{-3}$ to $\rho_B=0.0
- 0.588$ fm$^{-3}$ on account of the presence of hyperons. The
changes of the $^{1}S_{0}$ protonic SF can further suppress the NS's
cooling rate. And hyperons in NSs change the $^{1}S_{0}$ protonic
pairing gap which must affect the cooling properties of NSs.

Our model may simplify because it adopts the lowest level of
approximation in the BCS equation as well as neglects the possible
influence of the inhomogeneous in NS crust and $^{1}S_{0}$ hyperonic
pairing in NS core on the $^{1}S_{0}$ nucleonic energy gap. While it
can give the clear influence of hyperons degrees of freedom on the
$^{1}S_{0}$ nucleonic pairing. We will analyze more complicated
models in future studies.

\normalem
\begin{acknowledgements}
This work is funded by the National Natural Science Foundation of
China under grant Nos.11373047, 11103047 and 11303063.
\end{acknowledgements}

\bibliographystyle{raa}
\bibliography{bibtex}

\end{document}